# Towards Real-Time Respiratory Motion Prediction based on Long Short-Term Memory Neural Networks

Hui Lin[1], Chengyu Shi[2], Brian Wang[3], Maria F. Chan[2], Xiaoli Tang[2], Wei Ji[1*]

[1]Department of Mechanical Aerospace and Nuclear Engineering, Rensselaer Polytechnic Institute, Troy, NY, United States of America
[2]Department of Medical Physics, Memorial Sloan Kettering Cancer Center, New York, NY, United States of America
[3]Department of Radiation Oncology, University of Louisville, Louisville, KY, United States of America

*Corresponding Author E-mail: jiw2@rpi.edu

**Abstract**

Radiation therapy of thoracic and abdominal tumors requires incorporating the respiratory motion into treatments. To precisely account for the patient's respiratory motions and predict the respiratory signals, a generalized model for predictions of different types of patients' respiratory motions is desired. The aim of this study is to explore the feasibility of developing a Long Short-Term Memory (LSTM)-based generalized model for the respiratory signal prediction. To achieve that, 1703 sets of Real-Time Position Management data were collected from retrospective studies across three clinical institutions. These datasets were separated as the training, internal validity and external validity groups. Among all the datasets, 1187 datasets were used for model development and the remaining 516 datasets were used to test the model's generality power. Furthermore, an exhaustive grid search was implemented to find the optimal hyper-parameters of the LSTM model. The hyper-parameters are the number of LSTM layers, the number of hidden units, the optimizer, the learning rate, the number of epochs, and the length of time lags. The obtained model achieved superior accuracy over conventional artificial neural network models: with the prediction window equaling to 500ms, the LSTM model achieved an average relative Mean Absolute Error (MAE) of 0.037, an average Root Mean Square Error (RMSE) of 0.048, and a Maximum Error (ME) of 1.687 in the internal validity data, and an average relative MAE of 0.112, an average RMSE of 0.139 and an ME of 1.811 in the external validity data. Compared to the LSTM model trained with default hyper-parameters, the MAE of the optimized model results decreased by 20%, indicating the importance of tuning the hyper-parameters of LSTM models to obtain superior accuracies. This study demonstrates the potential of deep LSTM models for the respiratory signal prediction and illustrates the impacts of major hyper-parameters in LSTM models.

## 1. Introduction

The precision of radiotherapy at specific sites, such as lung, liver, breast, and pancreatic, can be greatly influenced by the respiratory motion of a patient. A modern radiotherapy system is expected to (1) detect and predict the patient-specific respiratory motions ahead of time and (2) accommodate the radiotherapy planning and delivery to the breathing-induced motion patterns. A number of motion management methods have been investigated: breathing-hold (Murphy *et al.*, 2002; Nelson *et al.*, 2005; Petralia *et al.*, 2012; Chen *et al.*, 2015) and gating method (Shirato *et al.*, 2000; Berbeco *et al.*, 2005; Yan *et al.*, 2005) are employed to reduce the treatment field and minimize the overdose to surrounding organs at risk (OAR); multileaf collimator (MLC)- and robotic couch-based (D D'Souza *et al.*, 2005; Jiang, 2006; Yi *et al.*, 2008; Sawant *et al.*, 2009; Han‐Oh *et al.*, 2010; Buzurovic *et al.*, 2011; Hansen *et al.*, 2016) dynamic tracking methods have been explored for real-time target tracking to improve the delivery efficiency. When respiratory motion is incorporated into the delivery strategy, target motion needs to be predicted ahead of time by a certain amount in order to compensate the latency of beam and field adjustments (Shepard *et al.*, 2018).

Existing prediction methods of respiratory motion fall into two major categories: model-based and learning-based. Model-based methods estimate predictions of respiratory motion by linear or non-linear functions and predictor coefficients. Existing model-based methods include Autoregressive Moving Average Model (ARIMA) (Hibon and Makridakis, 1997; McCall and Jeraj, 2007), Sinusoidal Model (Wu *et al.*, 2004; Vedam *et al.*, 2004) and Kalman Filter (Putra *et al.*, 2006; Wang and Balakrishnan, 2003). Learning-based methods employ neural networks or adaptive filters with higher complexity compared to the model-based methods, thus providing greater adaptability and nonlinearity to predict respiratory motion of patients, especially when the breathing pattern is irregular and fuzzy. Artificial Neural Network (ANN) and Recurrent Neural Network (RNN) are two commonly used learning-based methods. ANN usually consists of one input layer, two to five hidden layers and one output layer. The input layer takes in the respiratory curves and passes them to hidden layers, which are composed of neurons with adjustable weights and biases. These weights and biases are iteratively trained with the back propagation method until the threshold of the cost function is hit. The output layer finally outputs the prediction of future respiratory amplitudes. Previous studies (Tsai and Li, 2008; Isaksson *et al.*, 2005; Sun *et al.*, 2017) have demonstrated the superiority of ANN or its variant architectures in the breathing motion prediction especially for patients with nonlinear breathing curves. The major limitation of ANN is the ignorance of temporal dependency of previous inputs. RNN, on the other hand, is designed to process temporal data, and a feedback architecture is utilized to allow signals from previous hidden and output layers to feed back to current hidden and input layers. In this way, signals from very irregular breathing patterns can be supplemented by filtered data from the outputs and the response of the network becomes smooth. Lee *et al.* (Lee *et al.*, 2012) have explored the usage of RNN as the predictor of respiratory curves and encountered a problem regarding the expensive gradient descent calculations. As a result, they employed the Extend Kalman Filter (EKF) (Puskorius and Feldkamp, 1994) as the corrector of the RNN predictor. Although the architecture of RNN is shown to be a better fit for temporal data, in practice, it is hard to train it properly. The widely known issue is the exploding and vanishing gradient problem. Due to the long-term temporal contributions, the norm of the gradients either explodes to infinity or shrinks to zero, making it impossible to learn from timely distant events. As a result, a variant architecture of deep neural networks called Long Short-Term Memory (LSTM) (Hochreiter and Schmidhuber, 1997) was developed to address the problem of exploding and vanishing gradient in conventional RNN, achieving state-of-the-art performance in many sequential data tasks such as speech recognitions (Graves *et al.*, 2013; Graves and Jaitly, 2014) and machine translations (Cho *et al.*, 2014; Sutskever *et al.*, 2014).

In the previous studies of predictive model development, one generally used clinical data from limited number of patients to train the model. The models can provide accurate predictions for the same patients whose respiratory curve data were used to develop the model (Sun *et al.*, 2017; Teo *et al.*, 2018; Murphy and Pokhrel, 2009). However, limited results were presented when the model was applied to new patients, whose data were never used for model development. Large amount of clinical investigations have shown that the respiratory pattern can vary significantly from patient to patient (Chen and Kuo, 1989; Parameswaran *et al.*, 2006; Ozhasoglu and Murphy, 2002), and even for the same patient under different physical conditions (e.g. overweight or underweight, before or after surgery). Therefore, it will be more clinically practical if a generalized model is developed to predict respiratory signals for different patients. To develop such a generalized model, respiratory data from a large number of patients with considerably different breathing features are needed. Moreover, a powerful neural network that can "remember" and "resolve" tremendous amount of different data patterns needs to be constructed.

In this study, a novel LSTM-based approach was proposed to exploit recent developments in deep learning and its power in providing superior predictions of patient respiratory curves. The proposed deep learning neural network was trained using Real-Time Position Management (RPM) data from 1187 sets of respiratory signals, and the effectiveness of the network was first self-validated using the validation data that came from the same 1187 respiratory curves. This validation process is called *internal validity*, as the patterns of these validation data were learned during the training process, while the internal validity data remained untouched in the training process. Data sets from additional 516 breathing curves were then used as the *external validity* data to test the generality of the developed model. These data were not reached until the LSTM model had been trained and validated across the internal validity data. To our knowledge, this is the first time that a deep LSTM method has been developed as a generalized model for patient RPM predictions and with by far the largest amount of data employed for model training and validation (1703 RPM data from 985 patients in total).

**2. Methods and Materials**

*2.1. Data Acquisition*

The respiratory data (in total 1703 sets) were collected from 985 patients across three institutions by the Real-time Position Management (RPM; Varian Medical Systems, Palo Alto, CA) system. Patients were not subject to any active breathing motion management techniques (such as coaching or breath hold) during the data acquisition process, therefore the breathing curve can be considered as the record of each patient's free breathing pattern. The average frequency of the data is 30 Hz, and the recorded length ranged two to five minutes.

*2.2. Long Short-Term Memory networks*

Long Short-Term Memory (LSTM) is one of the most effective sequential models that can overcome the vanishing and exploding gradient problem during the long-term sequence learning compared to conventional RNN. The core idea of LSTM (Hochreiter and Schmidhuber, 1997) is to introduce a memory cell that enforces constant errors flowing through non-linear gating units and can truncate the gradient at some point. The decision of when to open the gating unit, close the gating unit, or delete the gating unit's access to the data is learned through an iterative process. The basic structure of an LSTM block is illustrated in Figure 1. The LSTM block can be simply decomposed into four major parts: a memory cell $s^t$, an input gate $i^t$, a forget gate $f^t$, and an output gate $o^t$. To train an LSTM model consisting of multiple LSTM blocks, at a specific time step $t$, the core step is to update the memory cell state $s^t$ by a self-loop. The control of the input, forget, and output gates of LSTM is achieved by specific weight and bias matrices. The details of updating the memory cell and each gate are illustrated as follows:

Let the input vector be denoted by $x^t$ and the current hidden layer vector be denoted by $h^t$. Vectors $b$, $U$, and $W$ denote the biases, input weights and recurrent weights respectively, and the subscripts *z*, *i*, *f* and *o* denote the block input, the input gate, the forget gate and the output gate. $\odot$ denotes the element-wise multiplication. The logistic sigmoid function $\sigma(x) = \frac{1}{1+e^{-x}}$ was employed as the activation function of gates, and the hyperbolic tangent function was usually used in the block input and block output.

The input gate $i^t$ is responsible for filtering the information (from the input vector and the previous hidden layer vector) that can be transferred to the memory cell. $i^t$ is updated by:

$$i^t = \sigma(U_i x^t + W_i h^{t-1} + b_i) \quad (1)$$

The forget gate $f^t$ controls the self-loop of the memory cell and decides which information from the previous memory cell state needs to be neglected. $f^t$ is updated by:

$$f^t = \sigma(U_f x^t + W_f h^{t-1} + b_f) \quad (2)$$

The memory cell controls the update of cell state from $s^{t-1}$ to $s^t$, and is updated with the use of block input $z^t$:

$$s^t = i^t \odot z^t + f^t \odot s^{t-1} \quad (3)$$

$$z^t = \tanh(U_z x^t + W_z h^{t-1} + b_z) \quad (4)$$

The output gate $o^t$ is responsible for generating the output and updating the current hidden layer vector $h^t$ with the use of block output $a^t$:

$$o^t = \sigma(U_o x^t + W_o h^{t-1} + b_o) \quad (5)$$

$$a^t = \tanh(s^{t-1}) \quad (6)$$

$$h^t = a^t \odot o^t \quad (7)$$

The predictive model was built with three LSTM layers, each with fifteen hidden layer units. The output layer was a fully connected layer with fifteen units, designed to output the predicted sequence of respiratory signals. The optimization of hyper-parameters such as the number of hidden layer units and the number of LSTM layers will be illustrated in Section 2.5.

*2.3. Data Partitioning*

As the starting point, the offset of each breathing curve was removed and the curves were normalized to -1 to 1. The general scheme of data partitioning is outlined in Figure 2. Each training dataset was separated into a training part and an internal validity part. The input of the LSTM model was denoted as $x_i$, a vector that contains $m$ data points and represents a segment of the breathing signal. The aim of training was to provide predictions of next $n$ data points right after $x_i$, denoted as $y_i$. The input $x_{i+1}$ was generated by forwarding $x_i$ with a fixed sliding window length, where the length was set to $n$ in order to continuously predict every data point right after the training input. We continued moving the sliding window until the last available observation in the training part was hit. Owing to the fact that the frequency of respiration signal is about 30 Hz, the length of $n=15$ corresponded to the prediction window of 500 ms. The input length $m$, also known as the length of time lags, is an important hyper-parameter in an LSTM architecture, so the value of $m$ was optimized. The details of this optimization are illustrated in Section 2.5.

*2.4. Model Training and Evaluation Process*

In the model training process, online learning was performed, which means the batch size was set to one and the network weights were updated after each training pattern. The $m$ data points within the training sample window were provided as LSTM inputs.

Predictions of the following *n* output data points were generated, and the errors were calculated. The sliding window was moved forward, and the weights and biases of the LSTM model were updated after each epoch. The model was trained for a fixed number of epochs, where a single epoch is a complete pass over the training set. After each epoch, we evaluate the model's Mean Absolute Error (MAE) using the output data in the training set. The network parameters for the next training epoch were inherited from the preceding training epoch, where the inherited parameters served as the initialization and regularization for the subsequent training epoch. The LSTM model was trained with the breathing signals in the training dataset, and the model evaluation was two-fold: first, the pre-trained model was evaluated using the internal validity data, where Mean Absolute Error (MAE), Root Mean Square Error (RMSE) and Maximum Error were used as the model performance metrics; and second, as an external validity dataset, an additional 516 sets of breathing signals, which were not ever used or "seen" through the model training process, were employed to assess the generality of the LSTM model.

*2.5. Hyper-parameter Optimization*

Hyper-parameter selection and optimization play an important role in obtaining state-of-the-art accuracy with LSTM networks (Greff *et al.*, 2017; Reimers and Gurevych, 2017). Although there are other methods (Bergstra and Bengio, 2012) that can efficiently obtain good hyper-parameters for complex networks, we chose to use the grid search method as it is easy to implement and parallelize. In the grid search, exhaustive searching was performed through the hyper-parameter space that consists of manually specified parameter subsets. Specifically, we performed 44,100 random searches, one for each combination of the six variants and each encompassing ten trials. Since the training for the LSTM model is time-consuming, the performance of the model with each parameter combination was evaluated by Mean Absolute Errors using a forward out-of-sample validation instead of cross-validation. For this LSTM network, we have investigated the following hyper-parameters.

**Number of LSTM layers**. Networks with a 1, 2, 3, 5 or 10 stacked LSTM layers were investigated.

**Number of hidden units per layer**. The number of hidden units per LSTM layer was selected from the set {3,5,10,15,20,30,40}. In case of multiple layers, we assigned the same value for each LSTM layer.

**Optimizer**. The optimizer is responsible to minimize the objective function of the LSTM networks. In this study we investigated the performance of seven optimizers, including stochastic gradient descent (SGD) (Robbins and Monro, 1985), Adam and Adamax (Kingma and Ba, 2014) and Nesterov Adam (Nadam) (Dozat, 2016), Adagrad (Duchi *et al.*, 2011), Adadelta (Zeiler, 2012) and RMSprop (Tieleman and Hinton, 2012).

**Learning rate**. The optimizer performance is affected by the learning rate. If the learning rate is too small, the training time will be very long; on the other hand, if the learning rate is too big, it may oscillate around the global optimum instead of converging to it. The learning rate was sampled from {0.0001,0.001,0.005,0.01,0.1}.

**Number of epochs**. The number of epochs is a hyper-parameter that defines the number of times that the LSTM networks passes through the entire training dataset. One epoch means that each sample in the training dataset has had an opportunity to update the internal model parameters. An epoch is comprised of one or more batches. The number of epochs is traditionally large, allowing the learning algorithm to run until the error from the model has been sufficiently minimized. In this study, the number of epochs was selected from the set {10,20,30,50,100,500}.

**Length of time lags**. Time lag refers to a sequence of the respiration signal acting as the LSTM model's input. The length of time lag represents the amount of intakes to make prediction, and different length of time lag may cause different prediction results. The length of time lag was sampled from the set {1,5,15,20,50,100}.

*2.6. Evaluation of Predictive Accuracy*

Mean Absolute Error (MAE), Root Mean Square Error (RMSE) and Maximum Error (ME) were used to evaluate the respiratory signal predictions generated by the LSTM model. The evaluation criteria indicated the overall prediction ability of the model by comparing (1) the internal validity dataset and (2) the external validity dataset with their corresponding predicted values.

MAE is a measure of the magnitude of errors, and can be calculated by

$$MAE = \frac{1}{N}\sum_{i=1}^{N}|y_i - \hat{y}_i| \qquad (8)$$

, where $y_i$ is the actual respiratory data point, $\hat{y}_i$ is the predicted respiratory data point, and N is the number of investigated points.

RMSE is calculated by taking the square root of the mean of the square of all the errors. The RMSE represents the sample standard deviation of the differences between predicted values and ground truth values. The effect of each error on RMSE is proportional to the size of the squared error. Consequently, RMSE is sensitive to outliers. RMSE can be expressed by:

$$RMSE = \sqrt{\frac{1}{N}\sum_{i=1}^{N}(y_i - \hat{y}_i)^2} \qquad (9)$$

ME represents the maximum error occurred in the predicted breathing sequence compared to the true sequence, and is calculated by:

$$ME = \max\{|\hat{y}_i|, i = 1, 2, ..., N\} \qquad (10)$$

where $y_{mean}$ and $\hat{y}_{mean}$ are the mean value of actual and predicted respiratory signals respectively.

*2.7. Experiment Details*

The LSTM model construction, training, and evaluation were implemented in the high-level neural networks API Keras version 2.0.4 (Chollet, 2015) in the Python 3.6 environment (Van Rossum and Drake Jr, 1995). TensorFlow 1.4 (Abadi *et al.*, 2016) was the backend. An Ubuntu server with Xeon E5-2697 CPU and one Nvidia Quadro P6000 (8GB RAM, 64GB memory) was used to perform all the calculations. The model training time was 25.6 hours.

## 3. Results

*3.1. Hyper-parameter Tuning in LSTM*

Figure 3 depicted the MAE (in a relative unit) as a function of the optimizers, the number of epochs, the learning rates, the number of hidden units, the length of time lags, and the number of LSTM layers. Table I summarizes the best hyper-parameter values and their importance levels affecting the MAE. Among them, the number of LSTM layers has the greatest impact on the model. The best performance was from a three-layer LSTM model and the worst was from a ten-layer model, causing up to 0.03 difference in relative MAE. In comparison, the type of optimizers and the learning rate demonstrated only a modest impact on the predictive performance of the model, where the MAE difference varied from 0.006 to 0.0075. The number of hidden units, the number of epochs, and the length of time lags showed little impact on the predictive accuracy, leading to small MAE differences from 0.00375 to 0.007.

*3.2. Predictive Performance Evaluation*

Figure 4 depicts six representative cases of ground truth versus predicted respiratory curves using the generalized LSTM model. For trajectories a, b, d, and e, no drastic prediction errors were observed. Most of the errors occurred near the peaks (transitions from inhale to exhale) and troughs (transitions from exhale to inhale) of respirations. Figure 5 provides a summary of the predictive performance on respiratory signals that are used for the model's internal validation and external validation. The MAE, RMSE, and ME of the internal validity dataset and external validity dataset are averaged and summarized in Table II. Results in Figure 5 and Table II indicate that the LSTM model can provide accurate predictions of respiratory signal of patients even if the dataset was not used to establish the parameters for the LSTM model. On the other hand, the model errors (MAE and RMSE) on the external validity dataset are about three times greater compared to the internal validity dataset. This is reasonable since the LSTM model captured the respiratory patterns of the internal validity dataset during the training process.

**4. Discussion**

*4.1. Impact of Hyper-parameters*

**Number of LSTM layers**: As expected, the number of LSTM layers played the key role in the network performance—the deeper the LSTM model went, the better the prediction performance, with the law of diminishing returns applied. In fact, the MAE of the five-layer model was only slightly better than the three-layer model, with the price of an over-fitting issue and slower convergence speed. Therefore, we decided to choose the three-layer structure. It can also be seen in Figure 3 that the MAE increased significantly when the number of LSTM layers was ten, which may be caused by the over-fitting of the model.

**Number of hidden units per layer**: As the number of LSTM layers determined the depth of an LSTM model, the width of an LSTM model was determined by the number of hidden units per layer. Finding the optimal number of hidden units was not straightforward due to its dependency on different inputs and the number of LSTM layers. The optimal value in our case was found to be 15. However, as Table I reveals, the number of hidden units had a small impact on the results. Adding or removing 10 hidden units from a three-layer LSTM model only changed the performance by roughly 2%.

**Optimizers and learning rates**: Learning rate is the most important hyper-parameter for the optimization process. As shown in Figure 3, the model performance was sensitive to the change of learning rates. We observed that the variances for SGD and Adagrad were much smaller in comparison to other optimizers. We also measured the time an optimizer took to converge. According to our measurements, Adam (optimization) converged first and required the least number of training epochs to obtain a good performance, while SGD took the longest time to converge.

**Length of time lags**: The decision of the length of time lags depends on the periodic characteristics of the training data. Using many lagged values can enhance the probability of capturing time-dependent features, while at the same time increasing the dimensionality of the convergence, leading to over-fitting as well as difficulties in training the model. The LSTM model performed poorly when using a smaller number of time lags compared to the case in which it was trained with a large number of time lags.

**Number of epochs**: It was observed that the MAE increased significantly beyond a certain number of epochs. This phenomenon may be due to the over-fitting when the model was trained with too many epochs. In that case, the LSTM model did not learn the data pattern but only memorized the data.

*4.2. Performance Comparison with Other Architectures*

The superiority of neural network architectures in internal and external respiratory motion predictions has been demonstrated in previous studies (Yun *et al.*, 2012; Sun *et* *al.*, 2017; Teo *et al.*, 2018; Wang *et al.*, 2018). To our knowledge, this study is the first to implement a multi-layer LSTM architecture for external respiratory signal prediction utilizing over 1000 patient datasets. We have illustrated the pipeline to build up the LSTM model and approaches to tune the hyper-parameters of the model. The advantage of optimizing the hyper-parameters has also been investigated. The results demonstrated that tuning the hyper-parameters led to approximately a 20% increase in predictive accuracy and up to more than 80% improvements for certain parameters, in comparison to utilizing the default hyper-parameters. Thus, our study proves that tuning the hyper-parameters is vitally important to obtain good results using a deep LSTM model for respiratory motion prediction. Compared to previous studies that mainly focused on investigating the power of artificial neural networks (Sun *et al.*, 2017; Teo *et al.*, 2018), our LSTM model allowed connections through time and provided a mechanism to feed the hidden layers from previous steps (long-term and short-term) as additional inputs of the next step. Our results display state-of-the-art performance in terms of the MAE and RMSE, both in the internal validity and external validity scenarios. As for the calculation speed, although the training of our deep LSTM model took over 20 hours to finish, it is a one-time effort. Once the model was trained, it took only *5 ms* to deploy the pre-trained model to make predictions per prediction window (500 ms), thus making it possible to perform the real-time respiratory signal predictions in the clinic.

*4.3. Limitations and Future Work*

There are some limitations of the current study. The first is due to the current method of hyper-parameter tuning. Although the grid search can mostly cover the search space, the uniformity of each hyper-parameter is limited and the exhaustive search process is very time-consuming. In addition, some hyper-parameters correlate with each other, and can result in different performances when optimized simultaneously rather than tuned individually. For future work, we will explore some emerging parameter optimization methods such as the random search (Bergstra and Bengio, 2012) and the Bayesian optimization (Snoek *et al.*, 2012) that can efficiently search through the parameter space and incorporate the parameter interactions.

The second limitation involves the unknown correlations between internal and external respiratory motion patterns. As expected, there are non-negligible differences between the external respiratory signals and internal tumor motions, thus to make our LSTM model clinically useful, we need to fill in the gap between external respiration prediction and internal tumor motion prediction. This can be achieved by developing another model to learn the correlations between external and internal respiratory motions, which is beyond the scope of this study.

**5. Conclusion**

A deep LSTM model was developed for the external respiration signal prediction for radiotherapy and has demonstrated the state-of-the-art performance compared to optimized conventional neural networks. A total of 1703 sets of breathing curves collected from 985 patients across three institutions were used for the model training and validation, and the internal and external validity results have shown that the developed model can be generalized and applied to predict respiratory motions for patients over a wide range of conditions. We are the first to demonstrate the feasibility of developing a single generalized model for respiratory prediction. In addition, a large-scale study on the impacts of LSTM hyper-parameters was investigated and reported, illustrating the necessity of hyper-parameter tuning to boost the model performance and providing insights on the relative importance of these hyper-parameters. This study has demonstrated the feasibility of utilizing the deep LSTM model for external respiratory signal predictions, which can be further extended to track the dynamic tumor motions during the treatment delivery in the future.

**Disclosure of Conflict of Interest**

The authors have no relevant conflicts of interest to disclose.

**Acknowledgments**

This research was funded in part through the NIH/NCI Cancer Center Support Grant P30 CA008748. This study was originated from an independent study with Dr. Wei Ji and accomplished by Dr. Hui Lin while completing her dissertation at RPI under the research support by Dr. X George Xu.

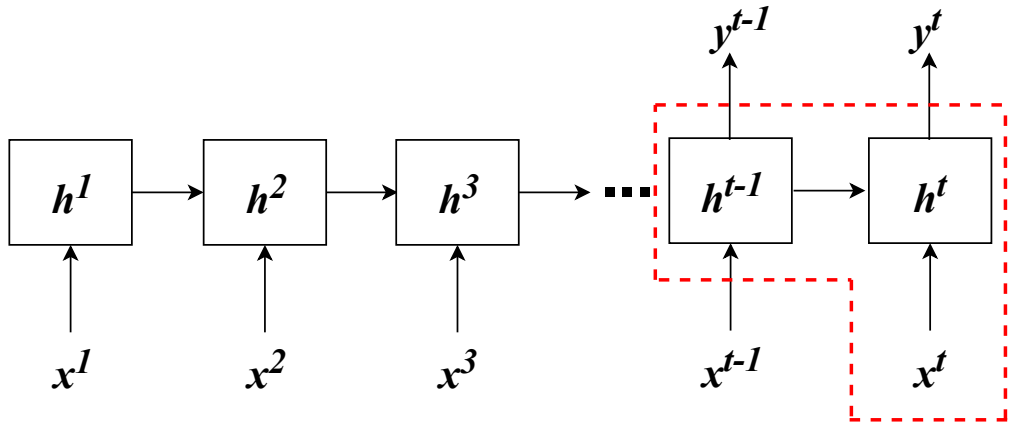

(a)

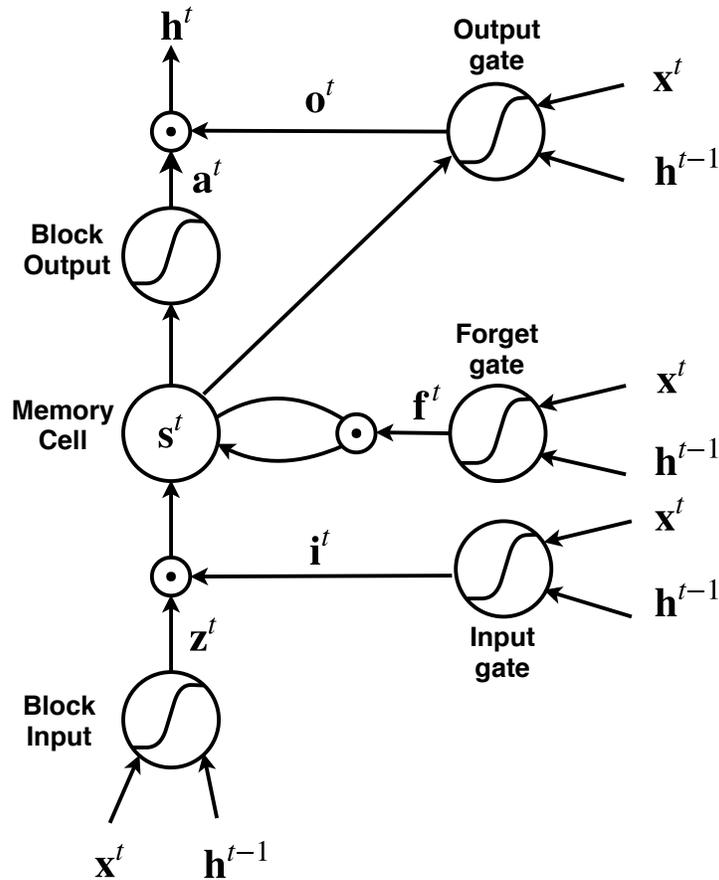

(b)

**Figure 1.** (a) The LSTM many-to-many architecture for respiratory sequence prediction. The input vector is denoted by $x^t$, the hidden layer vector is denoted by $h^t$, and the output vector is denoted by $y^t$. (b) A schematic diagram of a basic

LSTM block as delineated in (a) by the red dash lines. The LSTM block consists of four parts: a memory cell $s^t$, an input gate $i^t$, a forget gate $f^t$ and an output gate $o^t$. The updates of each part and the output of the LSTM block are illustrated in Equation (1) – (6).

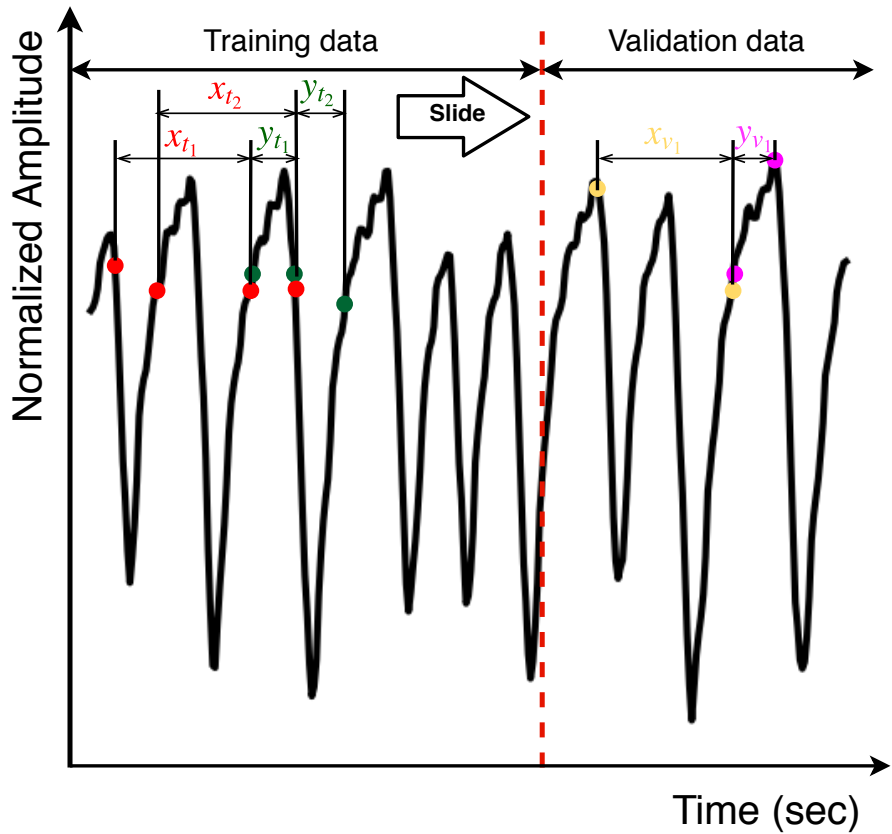

**Figure 2.** Data partitioning of the respiratory breathing signal. The breathing curves in the training dataset were first divided into the training part and the validation part. The training inputs ($x_{ti}$) and outputs ($y_{ti}$) of LSTM networks were generated via the sliding window technique. $x_{vi}$ and $y_{vi}$ represent the inputs and outputs of a pair set of internal validity data.

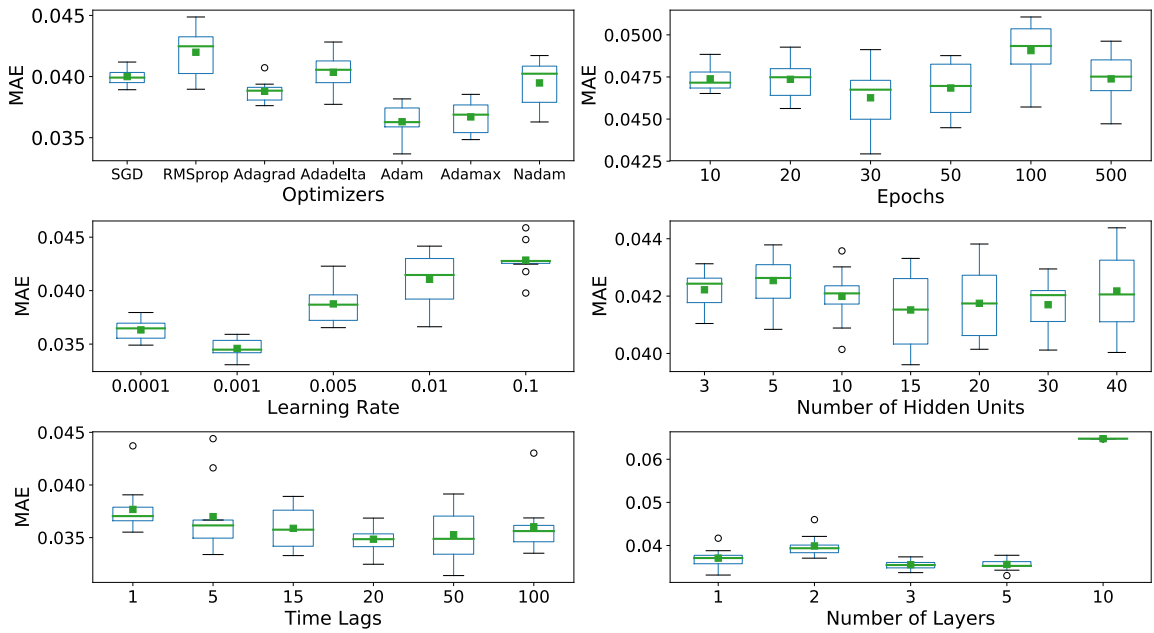

**Figure 3**. The MAE of predicted signals (mean value in green) with different values of different optimizers, the number of epochs, learning rates, the number of hidden units, lengths of time lags, and the number of LSTM layers. The MAEs were calculated in the internal validity dataset. The box plot indicates the standard deviation between the predicted marginal and thus the reliability of the predicted mean performance.

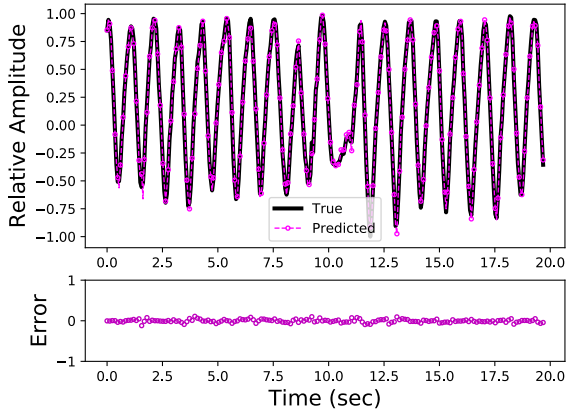

(a) Patient pattern 1: deep in depth and fast.

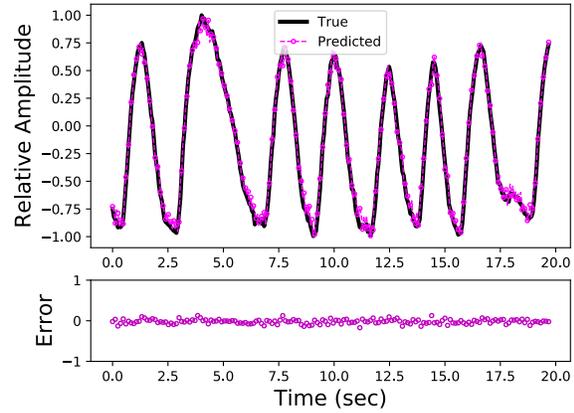

(b) Patient pattern 2: deep in depth and slow.

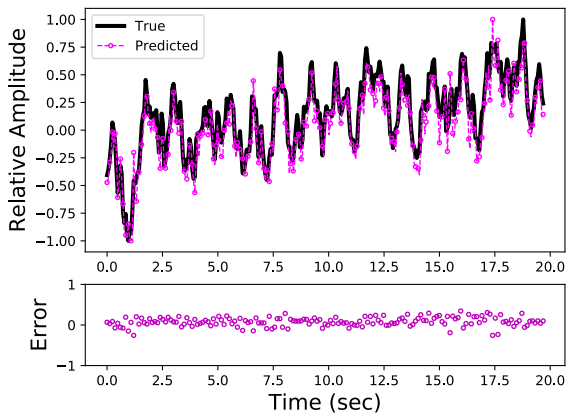

(c) Patient pattern 3: shallow in depth and fast.

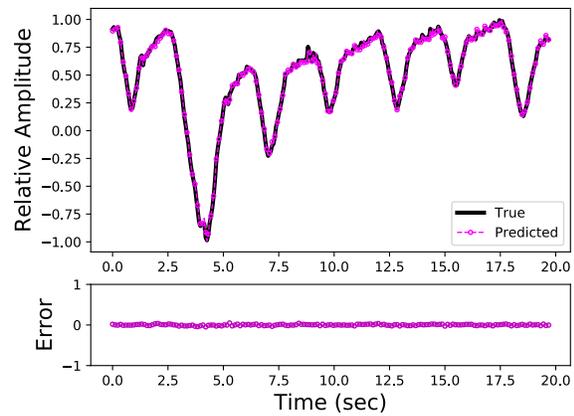

(d) Patient pattern 4: shallow in depth and slow.

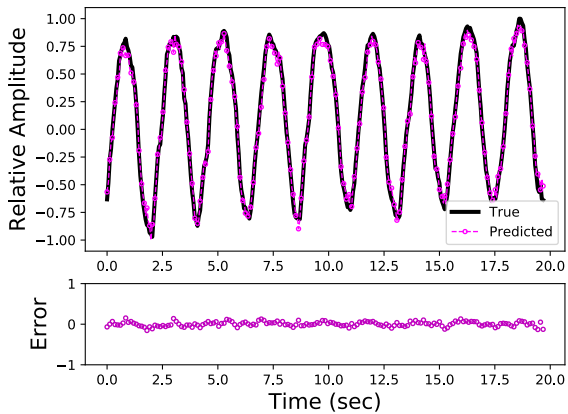

(e) Patient pattern 5: deep in depth and with a normal speed.

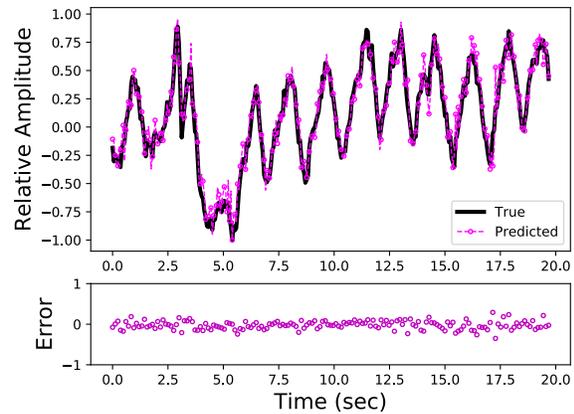

(f) Patient pattern 6: shallow in depth and with a normal speed.

**Figure 4.** Predicted respiratory signals benchmarked against the ground truth signals (selected from six patients with different breathing frequencies and amplitudes). The ground truth signals are depicted in black, and the predicted

signals are depicted in magenta. The absolute error plot is attached below each subplot.

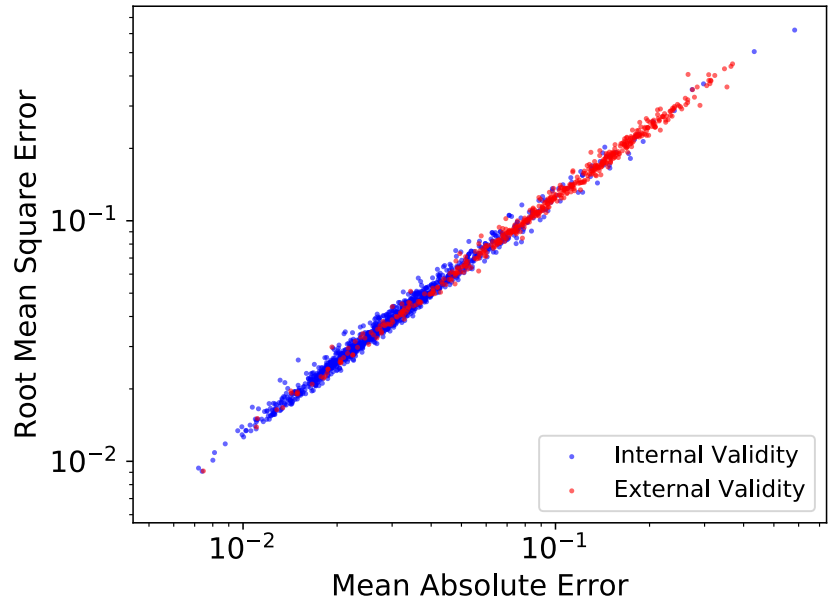

(a) Distribution of MAE and RMSE values for all patients.

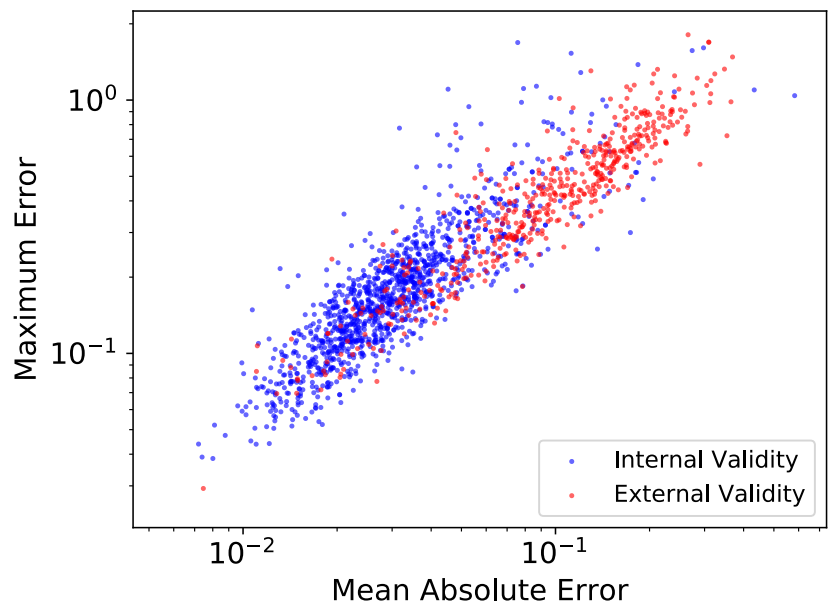

(b) Distribution of MAE and ME values for all patients.

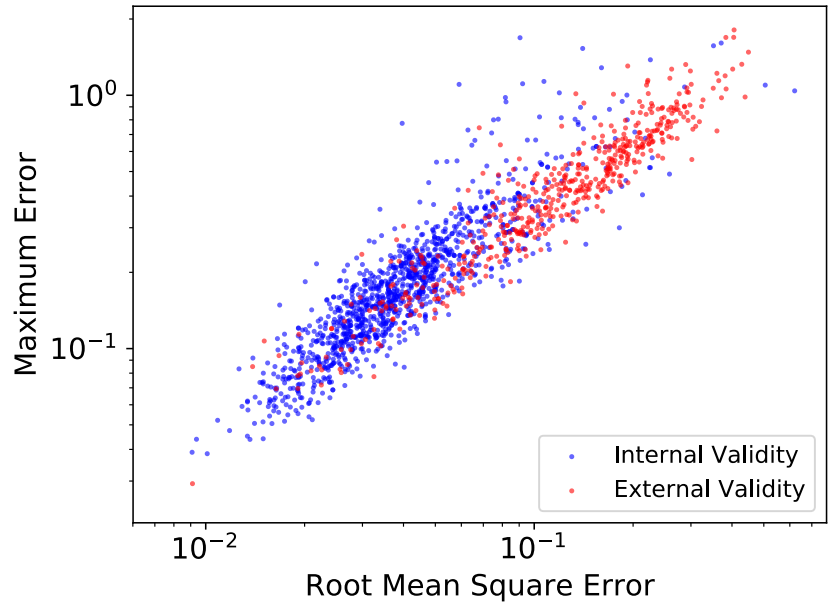

(c) Distribution of RMSE and ME values for all patients.

**Figure 5.** Illustration of the LSTM model's predictive performance on the internal validity dataset and the external validity dataset using Mean Absolute Errors (MAE), Root Mean Square Errors (RMSE) and Maximum Errors (ME) metrics. The internal validity data is represented by blue dots, and the external validity data was represented by red dots.

**Table I.** The summary of LSTM hyper-parameters investigated in this study, the recommended configurations, and the impact level of each parameter.

| Hyper-parameter | Range | Recommended configuration | Impact |
|---|---|---|---|
| Number of layers | {1,2,3,5,10} | 3 | High |
| Number of hidden units per layer | {3,5,10,15,20,30,40} | 15 | Low |
| Learning rate | {0.0001,0.001,0.005,0.01,0.1} | 0.001 | Intermediate |
| Number of epochs | {10,20,30,50,100,500} | 30 | Low |
| Length of time lags | {1,5,15,20,50,100} | 20 | Low |
| Optimizer | {SGD,RMSprop,Adagrad,Adadelta,Adam,Adamax,Nadam} | Adam | Intermediate |

**Table II.** Mean Absolute Errors (MAE), Root Mean Square Errors (RMSE) and Maximum Errors (ME) in predicting patient respiratory signals for the internal validity patient set (includes 1187 sets of breathing curves) and the external validity set (includes 516 sets of breathing curves).

|  | MAE | RMSE | ME |
|---|---|---|---|
| Internal validity data (1187 breathing curves) | 0.037 | 0.048 | 1.687 |
| External validity data (516 breathing curves) | 0.112 | 0.139 | 1.811 |

# References

Abadi M, Barham P, Chen J, Chen Z, Davis A, Dean J, Devin M, Ghemawat S, Irving G and Isard M *OSDI,2016),* vol. Series 16*)* pp 265-83

Berbeco R I, Nishioka S, Shirato H, Chen G T and Jiang S B 2005 Residual motion of lung tumours in gated radiotherapy with external respiratory surrogates *Physics in medicine and biology* **50** 3655

Bergstra J and Bengio Y 2012 Random search for hyper-parameter optimization *Journal of Machine Learning Research* **13** 281-305

Buzurovic I, Huang K, Yu Y and Podder T 2011 A robotic approach to 4D real-time tumor tracking for radiotherapy *Physics in medicine and biology* **56** 1299

Chen H and Kuo C-S 1989 Relationship between respiratory muscle function and age, sex, and other factors *Journal of Applied Physiology* **66** 943-8

Chen Z-G, Xu L, Zhang S-W, Huang Y and Pan R-H 2015 Lesion discrimination with breath-hold hepatic diffusion-weighted imaging: a meta-analysis *World Journal of Gastroenterology: WJG* **21** 1621

Cho K, Van Merriënboer B, Gulcehre C, Bahdanau D, Bougares F, Schwenk H and Bengio Y 2014 Learning phrase representations using RNN encoder-decoder for statistical machine translation *arXiv preprint arXiv:1406.1078*

Chollet F 2015 Keras.

D D'Souza W, Naqvi S A and Cedric X Y 2005 Real-time intra-fraction-motion tracking using the treatment couch: a feasibility study *Physics in medicine and biology* **50** 4021

Dozat T 2016 Incorporating nesterov momentum into adam

Duchi J, Hazan E and Singer Y 2011 Adaptive subgradient methods for online learning and stochastic optimization *Journal of Machine Learning Research* **12** 2121-59

Graves A and Jaitly N *Proceedings of the 31st International Conference on Machine Learning (ICML-14),2014),* vol. Series*)* pp 1764-72

Graves A, Mohamed A-r and Hinton G *Acoustics, speech and signal processing (icassp), 2013 ieee international conference on,2013),* vol. Series*)*: IEEE) pp 6645-9

Greff K, Srivastava R K, Koutník J, Steunebrink B R and Schmidhuber J 2017 LSTM: A search space odyssey *IEEE transactions on neural networks and learning systems*

Hansen R, Ravkilde T, Worm E S, Toftegaard J, Grau C, Macek K and Poulsen P R 2016 Electromagnetic guided couch and multileaf collimator tracking on a TrueBeam accelerator *Medical physics* **43** 2387-98

Han-Oh S, Yi B Y, Lerma F, Berman B L, Gui M and Yu C 2010 Verification of MLC based real-time tumor tracking using an electronic portal imaging device *Medical physics* **37** 2435-40

Hibon M and Makridakis S 1997 ARMA models and the Box–Jenkins methodology

Hochreiter S and Schmidhuber J 1997 Long short-term memory *Neural computation* **9** 1735-80

Isaksson M, Jalden J and Murphy M J 2005 On using an adaptive neural network to predict lung tumor motion during respiration for radiotherapy applications *Medical physics* **32** 3801-9

Jiang S B *Seminars in radiation oncology,2006),* vol. Series 16*)*: Elsevier) pp 239-48

Kingma D P and Ba J 2014 Adam: A method for stochastic optimization *arXiv preprint arXiv:1412.6980*

Lee S J, Motai Y and Murphy M 2012 Respiratory motion estimation with hybrid implementation of extended Kalman filter *IEEE Transactions on Industrial Electronics* **59** 4421-32

McCall K and Jeraj R 2007 Dual-component model of respiratory motion based on the periodic autoregressive moving average (periodic ARMA) method *Physics in medicine and biology* **52** 3455

Murphy M J, Martin D, Whyte R, Hai J, Ozhasoglu C and Le Q-T 2002 The effectiveness of breath-holding to stabilize lung and pancreas tumors during radiosurgery *International Journal of Radiation Oncology* Biology* Physics* **53** 475-82

Murphy M J and Pokhrel D 2009 Optimization of an adaptive neural network to predict breathing *Medical physics* **36** 40-7

Nelson C, Starkschall G, Balter P, Fitzpatrick M J, Antolak J A, Tolani N and Prado K 2005 Respiration-correlated treatment delivery using feedback-guided breath hold: A technical study *Medical physics* **32** 175-81

Ozhasoglu C and Murphy M J 2002 Issues in respiratory motion compensation during external-beam radiotherapy *International Journal of Radiation Oncology* Biology* Physics* **52** 1389-99

Parameswaran K, Todd D C and Soth M 2006 Altered respiratory physiology in obesity *Canadian respiratory journal* **13** 203-10

Petralia G, Summers P, Viotti S, Montefrancesco R, Raimondi S and Bellomi M 2012 Quantification of variability in breath-hold perfusion CT of hepatocellular carcinoma: a step toward clinical use *Radiology* **265** 448-56

Puskorius G V and Feldkamp L A 1994 Neurocontrol of nonlinear dynamical systems with Kalman filter trained recurrent networks *IEEE Transactions on neural networks* **5** 279-97

Putra D, Haas O, Mills J and Burnham K 2006 Prediction of tumour motion using interacting multiple model filter

Reimers N and Gurevych I 2017 Optimal hyperparameters for deep lstm-networks for sequence labeling tasks *arXiv preprint arXiv:1707.06799*

Robbins H and Monro S 1985 *Herbert Robbins Selected Papers*: Springer) pp 102-9

Sawant A, Smith R L, Venkat R B, Santanam L, Cho B, Poulsen P, Cattell H, Newell L J, Parikh P and Keall P J 2009 Toward submillimeter accuracy in the management of intrafraction motion: The integration of real-time internal position monitoring and multileaf collimator target tracking *International Journal of Radiation Oncology* Biology* Physics* **74** 575-82

Shepard A J, Matrosic C K, Radtke J L, Jupitz S A, Culberson W S and Bednarz B P 2018 Characterization of clinical linear accelerator triggering latency for motion management system development *Medical physics* **45** 4816-21

Shirato H, Shimizu S, Kunieda T, Kitamura K, van Herk M, Kagei K, Nishioka T, Hashimoto S, Fujita K and Aoyama H 2000 Physical aspects of a real-time tumor-tracking system for gated radiotherapy *International Journal of Radiation Oncology* Biology* Physics* **48** 1187-95

Snoek J, Larochelle H and Adams R P *Advances in neural information processing systems,2012), vol. Series)* pp 2951-9

Sun W, Jiang M, Ren L, Dang J, You T and Yin F 2017 Respiratory signal prediction based on adaptive boosting and multi-layer perceptron neural network *Physics in Medicine & Biology* **62** 6822

Sutskever I, Vinyals O and Le Q V *Advances in neural information processing systems,2014), vol. Series)* pp 3104-12

Teo T P, Ahmed S B, Kawalec P, Alayoubi N, Bruce N, Lyn E and Pistorius S 2018 Feasibility of predicting tumor motion using online data acquired during treatment and a generalized neural network optimized with offline patient tumor trajectories *Medical physics* **45** 830-45

Tieleman T and Hinton G 2012 Lecture 6.5-rmsprop: Divide the gradient by a running average of its recent magnitude *COURSERA: Neural networks for machine learning* **4** 26-31

Tsai T-I and Li D-C 2008 Approximate modeling for high order non-linear functions using small sample sets *Expert Systems with Applications* **34** 564-9

Van Rossum G and Drake Jr F L 1995 *Python tutorial*: Centrum voor Wiskunde en Informatica Amsterdam, The Netherlands)
Vedam S, Keall P, Docef A, Todor D, Kini V and Mohan R 2004 Predicting respiratory motion for four-dimensional radiotherapy *Medical physics* **31** 2274-83
Wang F and Balakrishnan V 2003 Robust steady-state filtering for systems with deterministic and stochastic uncertainties *IEEE Transactions on Signal Processing* **51** 2550-8
Wang R, Liang X, Zhu X and Xie Y 2018 A Feasibility of Respiration Prediction Based on Deep Bi-LSTM for Real-Time Tumor Tracking *IEEE Access* **6** 51262-8
Wu H, Sharp G C, Salzberg B, Kaeli D, Shirato H and Jiang S B 2004 A finite state model for respiratory motion analysis in image guided radiation therapy *Physics in Medicine and Biology* **49** 5357
Yan H, Yin F-F, Zhu G-P, Ajlouni M and Kim J H 2005 Adaptive prediction of internal target motion using external marker motion: a technical study *Physics in medicine and biology* **51** 31
Yi B Y, Han-Oh S, Lerma F, Berman B L and Yu C 2008 Real-time tumor tracking with preprogrammed dynamic multileaf-collimator motion and adaptive dose-rate regulation *Medical physics* **35** 3955-62
Yun J, Mackenzie M, Rathee S, Robinson D and Fallone B 2012 An artificial neural network (ANN)-based lung-tumor motion predictor for intrafractional MR tumor tracking *Medical physics* **39** 4423-33
Zeiler M D 2012 ADADELTA: an adaptive learning rate method *arXiv preprint arXiv:1212.5701*